\documentclass[aps,prd,notitlepage,nofootinbib,superscriptaddress,preprintnumbers,10pt,twocolumn]{revtex4-1}
\pdfoutput=1
\usepackage[latin9]{inputenc}
\usepackage{color}
\usepackage{bm}
\usepackage{amsmath}
\usepackage{amssymb}
\usepackage{graphicx}
\graphicspath{{images/}}
\usepackage[unicode=true,bookmarks=false,breaklinks=false,pdfborder={0 0 1},backref=false,colorlinks=true]{hyperref}
\usepackage{breakurl}
\usepackage{epstopdf}
\usepackage{array}
\usepackage{natbib}
\usepackage{empheq}
\usepackage{csquotes}

\usepackage{units}

\makeatletter

\def\Mpl{M_{\rm Pl}}
\makeatother

\begin{document}
\preprint{YITP-19-92, IPMU19-0139}

\title{Constant roll and primordial black holes}

\author{Hayato Motohashi}
\affiliation{Center for Gravitational Physics, Yukawa Institute for Theoretical Physics, Kyoto University, Kyoto 606-8502, Japan}

\author{Shinji Mukohyama}
\affiliation{Center for Gravitational Physics, Yukawa Institute for Theoretical Physics, Kyoto University, Kyoto 606-8502, Japan}
\affiliation{Kavli Institute for the Physics and Mathematics of the Universe (WPI), The University of Tokyo Institutes for Advanced Study, The University of Tokyo, Kashiwa, Chiba 277-8583, Japan}

\author{Michele Oliosi}
\affiliation{Center for Gravitational Physics, Yukawa Institute for Theoretical Physics, Kyoto University, Kyoto 606-8502, Japan}

\date{\today}

\begin{abstract}%%%%%%%%%%%%%%%%%%%%%%%%%%%%%%%%%%%%%%%% 
The constant-roll inflation with small positive value of the constant-roll parameter $\beta\equiv \frac{\ddot\phi}{H\dot\phi}={\rm const.}$ has been known to produce a slightly red-tilted curvature power spectrum compatible with the current observational constraints. 
In this work, we shed light on the constant-roll inflation with negative $\beta$
and investigate how a stage of constant-roll inflation may realize the growth in the primordial curvature power spectrum necessary to produce a peaked spectrum of primordial black hole abundance. 
We first review the behavior of constant-roll models in the range of parameters $-\frac32<\beta<0$, which allows for a constant-roll attractor stage generating a blue-tilted curvature power spectrum without superhorizon growth. As a concrete realization, we consider a potential with two slow-roll stages, separated by the constant-roll stage, in a way that satisfies the current constraints on the power spectrum and the primordial black hole abundance. 
The model can produce primordial black holes as all dark matter, LIGO-Virgo events, or OGLE microlensing events.
Due to the range of different scalar tilts allowed by the constant-roll potential, this construction is particularly robust and testable by future observations. 
\end{abstract}%%%%%%%%%%%%%%%%%%%%%%%%%%%%%%%%%%%%%%%% 

\maketitle

\section{Introduction}%%%%%%%%%%%%%%%%%%%%%%%%%%%%%%%%%%%%%%%% 

Primordial black holes (PBH) can be generated from the direct collapse of a local peak of primordial perturbations (from the tail of their distribution) at horizon reentry, if their amplitude is sufficiently large. 
Masses of PBHs vary depending on the time of reentry at primordial epoch e.g.\ during radiation domination, and hence the denomination ``primordial''. 
PBHs can fill a number of interesting roles (for a review see \cite{Carr:2009jm,Carr2016,Sasaki2018}), such as constituting all or a fraction of dark matter, or being the origin of the LIGO-Virgo binary black hole merger events \cite{Bird:2016dcv,Sasaki:2016jop} or OGLE 5-year microlensing events~\cite{Niikura2019}, for the latter of which a potential connection to ``Planet 9'' was also considered~\cite{Scholtz:2019csj}.

It was clarified that asking for a sharp increase in the power spectrum responsible for production of PBHs in canonical single-filed inflation requires to go beyond the standard slow-roll (SR) scenario where $\frac{\ddot\phi}{H\dot\phi}\approx 0$~\cite{Motohashi2017b}. A scenario that has been explored in this context is ultra-slow-roll (USR) \cite{Tsamis2003,Kinney2005}, which allows, with the relation $\frac{\ddot\phi}{H\dot\phi}= -3$ and through an extremely flat potential, for a scalar spectral tilt $n_s-1=3$. 
In addition to its intrinsic interest, USR has been found in \cite{Motohashi2017b,Germani2017} to be a good modelization of inflection-point-type potentials (notably \cite{Garcia-Bellido2017}), also believed to produce a spike in the scalar power spectrum. The transient stage of USR is characterized by a growing mode of curvature perturbation on superhorizon scales, which indeed allows for a steep growth of the power spectrum. However this is by no means a necessary condition for a peak on small scales.

In the present work, we concentrate instead on a broader class of models, the constant-roll (CR) inflation, which generalizes SR and USR, allows for red-/blue-tilted spectrum, and does not necessarily yield the growing superhorizon mode~\cite{Martin2012}. 
The constant-roll models possess exact solutions, for which the constant-roll condition, $\frac{\ddot\phi}{H\dot\phi}\equiv\beta={\rm const.}$ for canonical inflation, is exactly satisfied~\cite{Motohashi2014,Motohashi2017a,Motohashi2017,Motohashi:2019tyj}
(see also \cite{Barrow:1990nv,Barrow:1994nt,Contaldi:2003zv,Kofman:2007tr} for earlier attempts).
The canonical constant-roll inflation with parameter $\beta\approx 0.015$ generates red-tilted spectrum compatible with CMB constraints~\cite{Motohashi2014,Motohashi2017a} (see also \cite{GalvezGhersi:2018haa}).
On the other hand, the dynamics of the inflaton with a different range of the constant-roll parameter has been also extensively studied recently~\cite{Morse:2018kda,Gao:2019sbz,Lin2019}.

In this work, we focus on the constant-roll inflation with the parameter range $-\frac32 < \beta < 0$
and apply it to the generation of PBHs.
We show that it indeed offers an alternative to the transient USR stage as a way to increase the power spectrum rapidly, which in turn may help model the generation of PBHs. 
With this parameter range, we can obtain not only the tilt $n_s - 1 = 3$ of the scalar power spectrum in the USR limit\footnote{Note that $n_s - 1 = 3$ is obtained without considering other stages of inflation. Matching the USR potential to a SR pre-transient stage leads to the maximum tilt $n_s - 1 = 4$~\cite{Byrnes2018}. An even steeper tilt was found in \cite{Carrilho:2019oqg}, making use of a non-SR pre-transient stage.},
but also various blue tilts with $0 < n_s - 1 < 3$.
In contrast to USR models, we do not have superhorizon growth of curvature perturbations.
The possibility to adjust the tilt is also interesting in that it renders the PBH production more robust to other observations.  
Near-future observations will provide precise constraints on the primordial power spectrum on small scales, which will allow us to test the constant-roll scenario.

This paper is organised as follows. In Sec.~\ref{sec:cr}, we review the CR scenario and the properties of its primordial power spectra. In Sec.~\ref{sec:2pot}, we explore a three-stage potential, in particular considering constraints on the curvature power spectrum. In Sec.~\ref{sec:3pot} we show with a three-stage potential how a viable PBH abundance can be produced. Finally, in Sec.~\ref{sec:conclusion} we discuss our results.

\section{Blue-tilted spectrum from constant-roll}\label{sec:cr}%%%%%%%%%%%%%%%%%%%%%%%%%%%%%%%%%%%%%%%% 

In this section we review the canonical constant-roll inflation \cite{Motohashi2014}\footnote{Our definition of \textit{constant-roll} is the condition \eqref{eq:crcondition}, based on the original work \cite{Martin2012,Motohashi2014}.
Note that some variant of the CR condition has been considered in the literature, e.g.\ in \cite{Atal2019}
the case $\frac{\ddot \phi}{H\dot\phi} \sim {\rm const.} \lesssim -3$  was considered to describe a potential with a small barrier. Within the scope of the present work, this situation can indeed be understood as an approximate CR behavior. However, since the CR parameter space and the potential studied are rather different, the respective conclusions are a priori independent.} 
and the possibility of obtaining a blue-tilted spectrum.
We first review the constant-roll background dynamics in Sec.~\ref{sec:cr_bg}, in particular for the parameter region $-\frac32<\beta<0$, and then scalar and tensor perturbations in Sec.~\ref{sec:cr_pert}.

\subsection{Background dynamics}\label{sec:cr_bg}%%%%%%%%%%%%%%%%%%%%%

The dynamics of canonical single-field inflation is 
governed by
the Einstein equations
\begin{align}
3\Mpl^2 H^2 &= \frac{1}{2}\dot{\phi}^2 + V(\phi)\,,\\
-2\Mpl^2\dot{H} &= \dot{\phi}^2\,,
\end{align}
and the Klein-Gordon equation 
\begin{equation} \label{KGeq}
\ddot{\phi} + 3H\dot{\phi} + V_\phi = 0\,,
\end{equation} 
where $\Mpl\equiv (8\pi G)^{-1/2}$ is the reduced Planck mass, $a(t)$ is the scale factor of a flat Friedmann-Lema{\^i}tre-Robertson-Walker spacetime with the measure $ds^2 = -dt^2 + a^2(t)\delta_{ij}dx^idx^j$, $H\equiv \dot{a}/a$ is the Hubble rate of expansion, an upper dot denotes a derivative w.r.t.~$t$, and derivatives of the potential are denoted with subscripts, i.e.~$V_\phi \equiv\partial V/\partial\phi$.

In the standard slow-roll approximation one neglects $\ddot{\phi}$ in \eqref{KGeq}, whereas in the ultra-slow-roll inflation one sets $V_\phi=0$. 
Hence they correspond to requiring $\frac{\ddot{\phi}}{H\dot{\phi}}\approx 0$ or $\frac{\ddot{\phi}}{H\dot{\phi}}=-3$, respectively.
The constant-roll inflation generalizes these cases, and is defined by a constant rate of roll
\begin{equation} 
\beta \equiv \frac{\ddot{\phi}}{H\dot{\phi}}\overset{!}{=} {\rm const}.\,, \label{eq:crcondition}
\end{equation}
where $\beta$ is a constant parameter characterising the model. (In other works on CR inflation other notations have been used, notably $\alpha \equiv - (3+\beta)$ in \cite{Motohashi2014} and $\eta \equiv -\beta$ in \cite{Lin2019}.)
In particular, the SR limit corresponds to $\beta\to 0$, and the USR limit is $\beta\to -3$.
Depending on the value of $\beta$, the \textit{CR condition} \eqref{eq:crcondition} leads to several sets of potential and background evolution, which can be obtained as exact solutions of the system of differential equations by using the Hamiltonian-Jacobi formalism. 
A $\cos$-type potential with $\beta\approx 0.015$ generates a slightly red-tilted spectrum with a small tensor-to-scalar ratio compatible with the latest observational constraints~\cite{Motohashi2014,Motohashi2017a}.
On the other hand, here we focus on a $\cosh$-type potential for $\beta<0$ (see Eq.~(19) of \cite{Motohashi2014})
i.e.\
\begin{equation} 
V(\phi) = 3 M^2\Mpl^2\left\{1-\frac{3 + \beta}{6}\left[1- \cosh\left( \sqrt{2|\beta|}\frac{\phi}{\Mpl}\right)\right]\right\},\label{eq:cr_pot}
\end{equation}
for which the exact solution for the background evolution is given by 
\begin{align} 
\phi &= \Mpl \sqrt{\frac{2}{|\beta|}}\ln\left[\coth\left(\frac{|\beta|}{2}M t\right)\right],\label{eq:cr_analytical} \\
H &= M \coth (|\beta| Mt) ,\label{eq:cr_h}
\end{align}
where $M$ is the energy scale of inflation. 
Out of all possibilities, we focus here on the range $-\frac{3}{2} < \beta < 0$, which allows for frozen super-Hubble curvature modes and a blue-tilted power spectrum, as reviewed further on.

\begin{figure}[t]
\centering
\includegraphics[width=.45\textwidth,trim={0 0 0 -1cm},angle=0,clip]{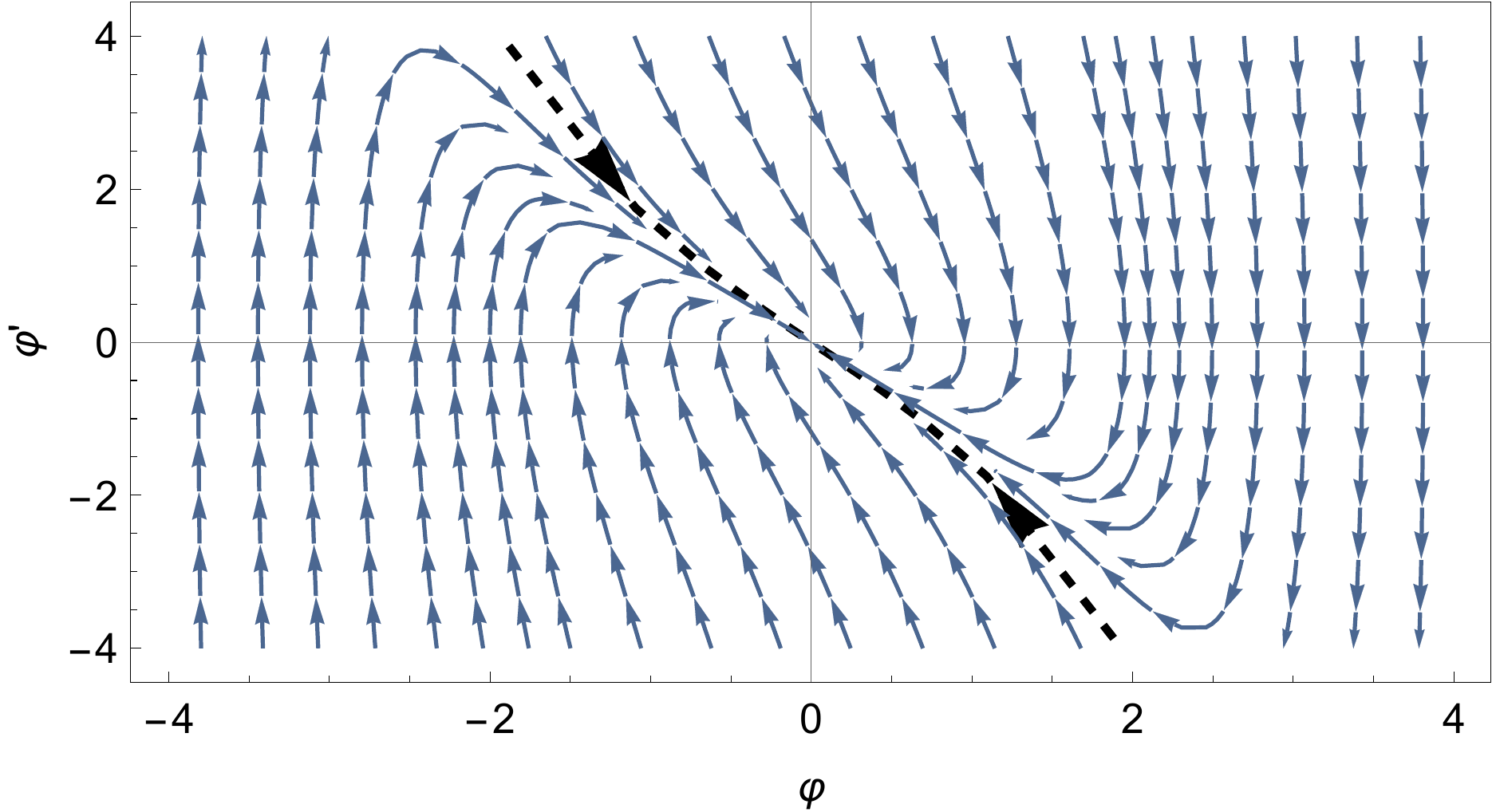}
\caption{Phase space evolution of the inflaton for the potential~\eqref{eq:cr_pot} with $\beta = -1.4$, 
where $\varphi \equiv \phi/\Mpl$ and $\varphi'\equiv d\varphi/dN$.
The analytical CR solution~\eqref{eq:cr_analytical} (\textit{dashed black}) is the attractor solution since in this case $-\frac{3}{2}<\beta < 0$. 
}
\label{fig:stage_alpM1P6}
\end{figure}

While the CR potential~\eqref{eq:cr_pot} allows the exact solution~\eqref{eq:cr_analytical},
the background evolution of the inflationary system for various initial conditions has to be checked numerically. For convenience in the numerical integration, we use dimensionless variables $\varphi \equiv \phi/\Mpl$ and $h \equiv H/M$, and the $e$-folding number $N\equiv \ln\frac{a}{a_i}$. 
Solving for a range of initial conditions $\{\varphi(0),\varphi'(0)\}$ with a prime denoting the derivative w.r.t.\ $N$, we obtain the phase space trajectories that characterize the background dynamics. 
It was clarified in \cite{Lin2019} that for the CR potential for a given value of $\beta<-3/2$ the numerical integration shows that the value of $\frac{\ddot\phi}{H\dot\phi}$ approaches to $-(3+\beta)$, and hence in that case (including USR with $\beta=-3$) the CR analytical solution~\eqref{eq:cr_analytical} is not the attractor solution. 
In the present case,
with $-\frac{3}{2}<\beta < 0$,
the CR analytical solution~\eqref{eq:cr_analytical} is the attractor of the system,
which can then be simply used to deduce the behavior of other quantities of interest. To illustrate the dynamics for the parameter range we are interested in, we depict in Fig.~\ref{fig:stage_alpM1P6} the phase space trajectories for $\beta = - 1.4$ and the corresponding CR analytical solution.

% ==================== Figure ====================
% ==================== Figure ====================

For further convenience, and in order to further characterize the dynamics, one may estimate using the analytical solution \eqref{eq:cr_h} the slow-roll parameters in the Hubble hierarchy $\epsilon_n\, (n = 1,2,3,\cdots)$, defined as customary 
\begin{equation}
\epsilon_1\equiv -\frac{\dot H}{H^2}\,,\quad \epsilon_{n+1}\equiv \frac{\dot{\epsilon}_n}{H\epsilon_n}\,.
\end{equation}
While $|\epsilon_n|\ll 1$ for the SR case, for the CR solution~\eqref{eq:cr_h} for $\beta<0$, these parameters approach (in the limit $N\gg1$) \cite{Motohashi2014}
\begin{equation} \label{srpara}
2\epsilon_1 = \epsilon_{2n+1} \simeq -2\beta a^{2\beta}\,,\quad \epsilon_{2n}\simeq 2\beta\,.
\end{equation}
In particular, $\epsilon_{2n}$ may take a non-negligible, asymptotically constant value, while the odd-indexed parameters, instead, quickly become very small. 
Hence, the asymptotic behavior is $2\epsilon_1 = \epsilon_{2n+1} \to 0$ and $\epsilon_{2n}\to 2\beta$, which is actually the same as those at the limit $Mt\to -\infty$ of the CR model for $\beta>0$ with the $\cos$-type potential.

In general, the production of PBHs as DM in canonical single-field inflation requires the no go of slow roll~\cite{Motohashi2017b}
\begin{equation} 
-\frac{\Delta \ln \epsilon_1}{\Delta N} > \mathcal{O}(1) .
\end{equation}
Using the slow-roll parameters~\eqref{srpara}, this condition reads $-2\beta>\mathcal{O}(1)$.
The USR $\beta= -3$ is a particular example that satisfies the condition.
More generally, it is actually possible for the CR scenario with general $\beta$ to satisfy this requirement.

The previous results allow us to estimate 
\begin{align}
\frac{1}{z}\frac{d^2z}{d\tau^2} &= a^2 H^2\left(2-\epsilon_1 + \frac{3}{2}\epsilon_2 + \frac{1}{4}\epsilon_2^2 - \frac{1}{2}\epsilon_1\epsilon_2 + \frac{1}{2}\epsilon_2\epsilon_3\right)\notag\\
&\simeq \frac{(1+\beta)(2+\beta)}{\tau^2} \equiv\frac{\nu^2-1/4}{\tau^2} \,,\label{eq:ms_bg}
\end{align}
where $\tau$ is the conformal time with $dt = a\, d\tau$, $z\equiv a \sqrt{\epsilon_1}$, and $\nu\equiv|\beta+3/2|$.
Here, the first equality exactly holds without any approximation, and the $\simeq$ sign holds when $\epsilon_1$ and $\epsilon_3$ are negligible and $\epsilon_2 \simeq 2\beta$. 
This result is useful for characterizing the perturbation equations (see Sec.~\ref{sec:cr_pert}).

\begin{figure}[t]
\centering
\includegraphics[width=8.5cm,angle=0,clip]{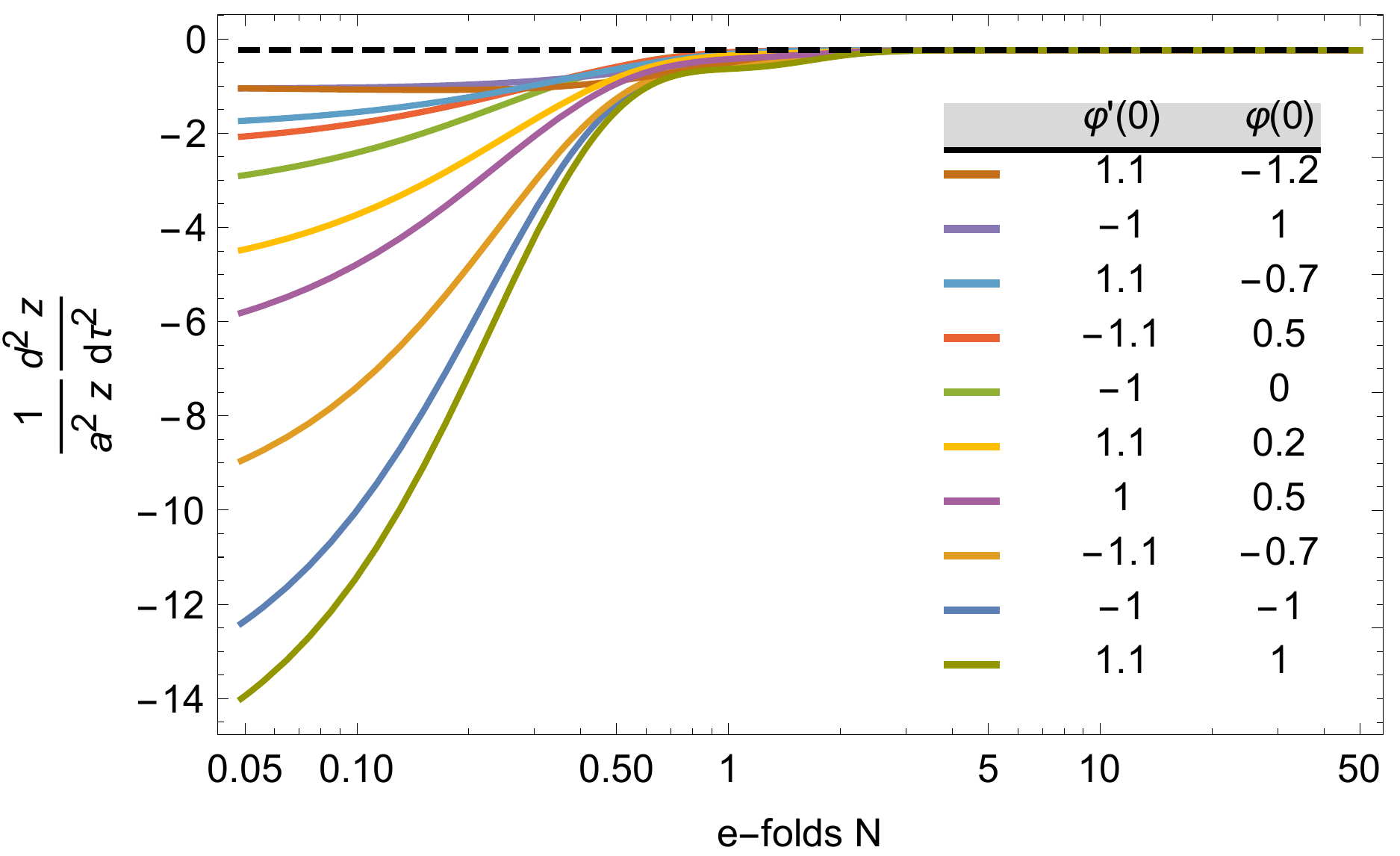}
\caption{Evolution of the background contribution \eqref{eq:ms_bg} to the Mukhanov-Sasaki equation \eqref{eq:ms} for $\beta = -1.4$ with various sets of initial conditions. 
All converge within \unit[$\mathcal{O}(1)$]{$e$-folds} towards the analytical solution.
}
\label{fig:zddzatr_alpM1P6}
\end{figure}

As a preparation for the numerical study of perturbations, one may also ask how well the CR analytical solution approximates a generic field-space trajectory. To answer this question we check the evolution of $\frac{1}{a^2 z}\frac{d^2z}{d\tau^2}$ for different initial conditions, and find that the analytical solution supports a wide basin of attraction (see Fig.~\ref{fig:zddzatr_alpM1P6}) and that it is reached within $\mathcal{O}(1)$ $e$-folds.
Therefore, even if one starts from initial conditions that are away from the attractor solution, after several $e$-folds, the analytical solution describes the system very well.

% ==================== Figure ====================
% ==================== Figure ====================

\subsection{Primordial power spectra}\label{sec:cr_pert}%%%%%%%%%%%%%%%%%%%%%

The curvature perturbations in Fourier space follow the Mukhanov-Sasaki (MS) equation
\begin{equation}
\frac{d^2v_k}{d\tau^2} + \left(\frac{1}{z}\frac{d^2z}{d\tau^2} + k^2\right)v_k = 0\,,
\label{eq:ms}
\end{equation}
where 
the first term in the parenthesis was determined in \eqref{eq:ms_bg}. The friction term is removed by using the MS variable $v_k \equiv \sqrt{2}\Mpl z\, \zeta_k$, instead of directly using the curvature perturbation $\zeta_k$ in the comoving gauge. When working in dimensionless variables, it is convenient to use $\tilde{k} \equiv k/M$. Finally, the initial condition for each mode function is taken to be the Bunch-Davies one, i.e.
\begin{equation}
{v_k}\bigr|_{k\tau\to-\infty} = \frac{1}{\sqrt{2k}}e^{-ik\tau}\,.
\end{equation}
Using Eq.~\eqref{eq:ms_bg}, which defines the constant $\nu$, one may predict the scalar curvature power spectrum. 
As confirmed in \cite{Martin2012,Motohashi2014}, for $\beta>-3/2$ the superhorizon solution of curvature perturbations consists of a constant and a decaying mode, and there is no superhorizon evolution in contrast to the case of USR. 
Hence, it is possible to maintain a CR stage for a long time without incurring an excessive growth of perturbations. Given the standard form of \eqref{eq:ms_bg}, the mode functions are given by the Hankel functions of the first type, i.e.\ \cite{Motohashi2014}
\begin{equation}
v_k(\tau) = \frac{\sqrt{-\pi\tau}}{2}H_\nu^{(1)}(-k\tau)\,.
\end{equation}
Then the scalar power spectrum is given by the standard result
\begin{equation}
\Delta^2_\zeta(k)\equiv \frac{k^3}{2\pi^2}|\zeta_k|^2 = \frac{H^2}{8\pi^2\Mpl^2\epsilon_1}\frac{2^{2\nu-1}\Gamma(\nu)^2}{\pi}\left(\frac{k}{aH}\right)^{3-2\nu}\,, \label{Ds2}
\end{equation}
where $|\zeta_k|$ is evaluated at the horizon exit assuming that it is frozen on superhorizon scales. From this expression one can predict the tilt of the scalar power spectrum as
\begin{equation}
n_s-1= 3-2\nu= 3-|2\beta+3|\,. \label{nsm1}
\end{equation}
Therefore, naively it would imply that the power spectrum is scale-invariant for $\beta=0$ or $-3$, i.e.\ SR or USR, 
red tilted for $\beta<-3$ or $\beta>0$, 
and blue tilted for $-3<\beta<0$.

However, a caveat is that the power spectrum~\eqref{Ds2} is evaluated at the horizon exit. 
For $\beta<-3/2$ the curvature perturbation actually grows on superhorizon scales until the end of the CR stage, which should be taken into account for the evaluation of the amplitude and tilt of the final power spectrum, in addition to the above calculation evaluated at the horizon exit.
As emphasized above, it is also important to note that the CR (or USR for the limit $\beta\to -3$) solution for $\beta<-3/2$ is not the attractor solution. 

On the other hand, the CR models with $\beta>-3/2$ are closer to the standard one since it does not have the above subtleties.
From \eqref{nsm1} we see that $\beta\approx 0.015$ can indeed explain the observed red-tilted spectrum $n_s\approx 0.97$~\cite{Motohashi2014,Motohashi2017a}.
Here, instead, we focus on a possibility for the CR stage to produce a blue-tilted spectrum on small scales, and hence focus on the parameter range $-3/2<\beta<0$, 
for which the CR solution is the attractor and the spectrum is blue with the tilt $0<n_s-1<3$ with curvature perturbation frozen on superhorizon scales.
The adjustable blue tilt in the CR inflation is useful for PBH production within the framework of canonical single-field inflation, which shall be discussed in more detail in Secs.~\ref{sec:2pot} and \ref{sec:3pot} below.

One may repeat the previous steps for tensor perturbations, 
yielding a scale-invariant spectrum \cite{Motohashi2014}
\begin{equation}
\Delta_t^2(k) = \frac{2H^2}{\pi^2\Mpl^2}\,.
\end{equation}

\section{Constructing a multi-stage potential}\label{sec:2pot}%%%%%%%%%%%%%%%%%%%%%%%%%%%%%%%%%%%%%%%% 

Building upon the results reviewed in Sec.~\ref{sec:cr}, in this section we construct a potential that transits from a standard SR potential, which we here choose as the Starobinsky inflation \cite{Starobinsky1980}, which is favored by the Planck CMB data, to the CR potential~\eqref{eq:cr_pot} responsible for producing a blue tilt, and then to the final SR stage that impedes the overproduction of PBHs and leads to end of inflation. 
The first SR stage allows to satisfy the CMB constraints~\footnote{Note that alternatively one may consider a two-field model such as a curvaton scenario \cite{Linde1996,Lyth2001,Moroi2001}, in which the curvaton would dominate the perturbation spectrum on CMB scales, and the CR field produces blue tilted spectrum on small scales. In this work we do not explore further this simple alternative.} while having a chance to produce PBHs from the CR stage. The Starobinsky potential in the Einstein frame is given by 
\begin{equation}
V_\textrm{SR1}(\varphi) =
m^2\Mpl^2\left(1 - e^{-\sqrt{2/3}(\varphi-\varphi_s)}\right)^2\,,
\end{equation}
where $m$ is the mass scale of the Starobinsky stage of inflation. 
Here we shifted the potential by a constant $\varphi_s$ to adjust the transition to the CR stage. 
Since we are mostly interested in the peak value of the power spectrum generated during the CR stage as well as in the typical behavior at the transition, and less in what occurs during the later stage of inflation, we simply approximate the second SR stage by a linear potential
\begin{equation} \label{VSR2}
V_\textrm{SR2}(\varphi) \simeq W_{\textrm{SR2}}\,\varphi + \Lambda_\textrm{SR2} \,, 
\end{equation}
with $W_{\textrm{SR2}},\,\Lambda_\textrm{SR2} = {\rm const}$.  For a more realistic model one should instead implement a potential that smoothly transitions to a stage of reheating, which is however beyond the scope of the present paper. Naturally, this latter stage should also ensure that the total number of $e$-folds of inflation is sufficient (i.e.~$N_\textrm{tot}\gtrsim50$).

Finally, as a phenomenological model for the transitions, we employ a $\tanh$-type smooth step-function,
\begin{equation}
\tilde{\Theta}_d
(x) \equiv \frac{1}{2} \left[ 1 + \tanh \left(\frac{x}{d}\right) \right],
\end{equation}
with an adjustable width $d$.

\begin{figure}
\includegraphics[width=8.5cm,angle=0,clip]{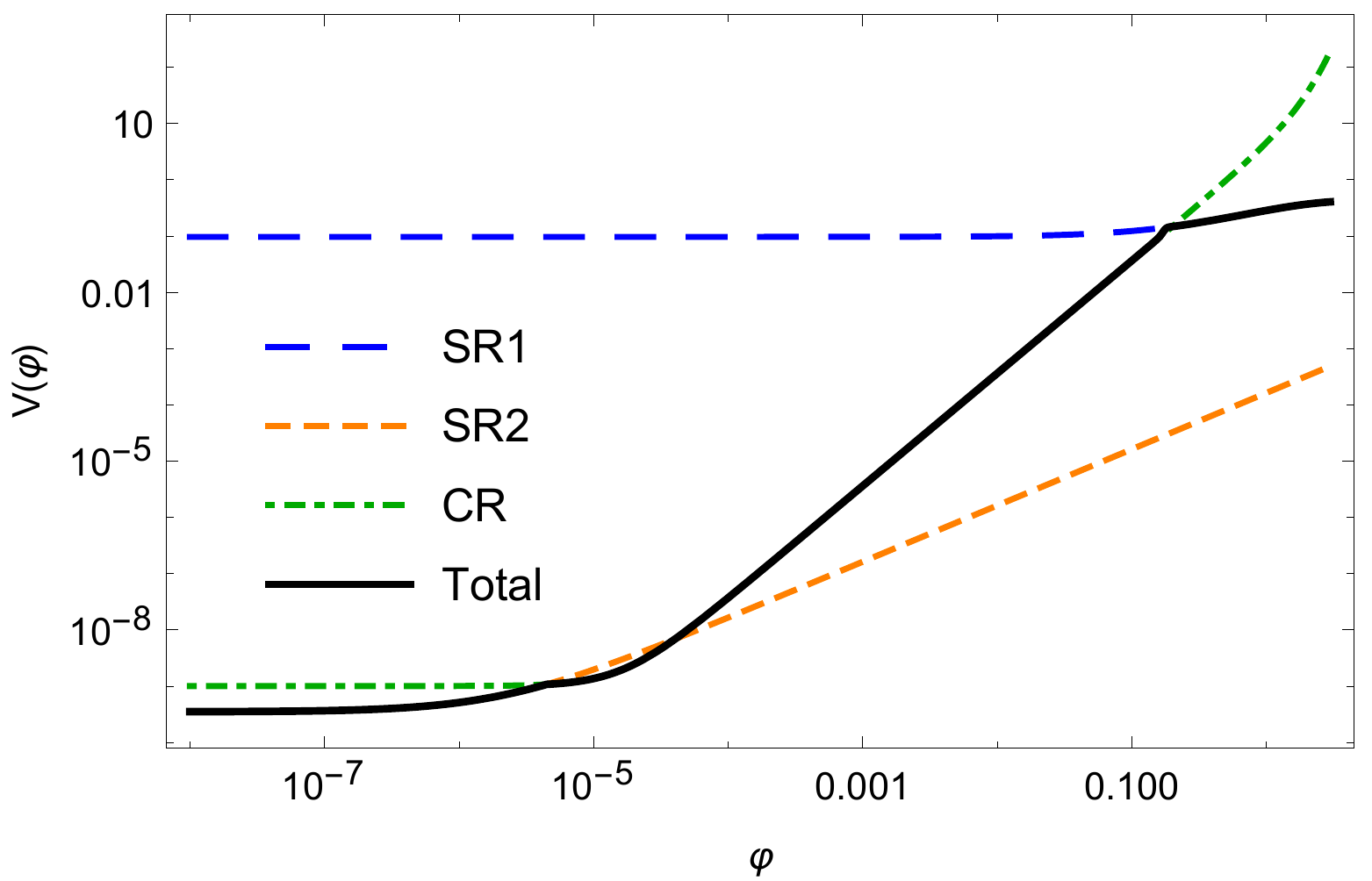}
\caption{The three-stage potential \eqref{eq:pot3step} with its different contributions, leading to the first SR stage for $\varphi \gtrsim \varphi_1$, then the CR stage for $\varphi_1 \lesssim \varphi \lesssim \varphi_2$, and finally the second SR stage for $\varphi \lesssim \varphi_2$. The parameters are chosen as in \eqref{eq:3step_parameters}.}
\label{fig:pot}
\end{figure}

The previous considerations lead ultimately to the full potential
\begin{align}
V(\varphi) &\equiv V_{\textrm{CR}}(\varphi)\tilde{\Theta}_{d_1}
(\varphi_1 - \varphi)\tilde{\Theta}_{d_2}
(\varphi-\varphi_2) \notag\\
&~~~ + V_\textrm{SR1}(\varphi)\tilde{\Theta}_{d_1}
(\varphi-\varphi_1) \notag\\
&~~~ +V_\textrm{SR2}(\varphi) \tilde{\Theta}_{d_2}
(\varphi_2 - \varphi)\,,\label{eq:pot3step}
\end{align}
with $\varphi_1 > \varphi_2$, where $d_1$ and $d_2$ are two constants determining the width of the first and second transitions. 
The potential $V_\textrm{CR}(\varphi)$ denotes the CR potential~\eqref{eq:cr_pot} expressed as a function of $\varphi \equiv \phi/\Mpl$. The potential \eqref{eq:pot3step} for the parameter set~\eqref{eq:3step_parameters} below is represented in Fig.~\ref{fig:pot}.

% ==================== Figure ====================
% ==================== Figure ====================

We set initial conditions and adjust the transition period to make the scalar and tensor power spectra generated during the first SR stage compatible with observational constraints. In order to obtain 
the first transition
approximately at the intended $\varphi_1$, we adjust $\varphi_s$ and the ratio $m/M$. In particular, it is convenient to minimize the difference in the slope of the potential between the first SR and CR stages. One may also adapt the width of the transition, which we keep relatively small in this work.

To fix the details of the second transition from CR to SR, one needs to consider the PBH distribution function. The second transition should indeed let the power spectrum decrease shortly after reaching the threshold amplitude for sufficient PBH production. 
This guarantees that
the PBH mass distribution is peaked enough to satisfy the thin windows allowed by current constraints, and also that the curvature power spectrum satisfies constraints on smaller scales. Leaving the details of the PBH distribution to Sec.~\ref{sec:3pot}, we specify here how to fix the parameters in the potential: 
The two constants $W_{\textrm{SR2}}$ and $\Lambda_{\textrm{SR2}}$ are fixed in such a way that the potential is continuous at $\varphi=\varphi_2$ in the limit $d_2\to 0$, i.e.~$V_\textrm{CR} (\varphi_2) = W_{\textrm{SR2}} \varphi_2 + \Lambda_{\textrm{SR2}}$ yet that the slope of the potential increases enough to guarantee a drop in the power spectrum, i.e.~$V_\textrm{CR}' (\varphi_2) < W_{\textrm{SR2}}$; $d_2$ is fixed such that the second transition is short enough (since the field value becomes exponentially small in $N$ during CR). 

\begin{figure}[t]
\centering
\includegraphics[width=10cm,angle=0,trim={.77cm 0 .3cm 0},clip]{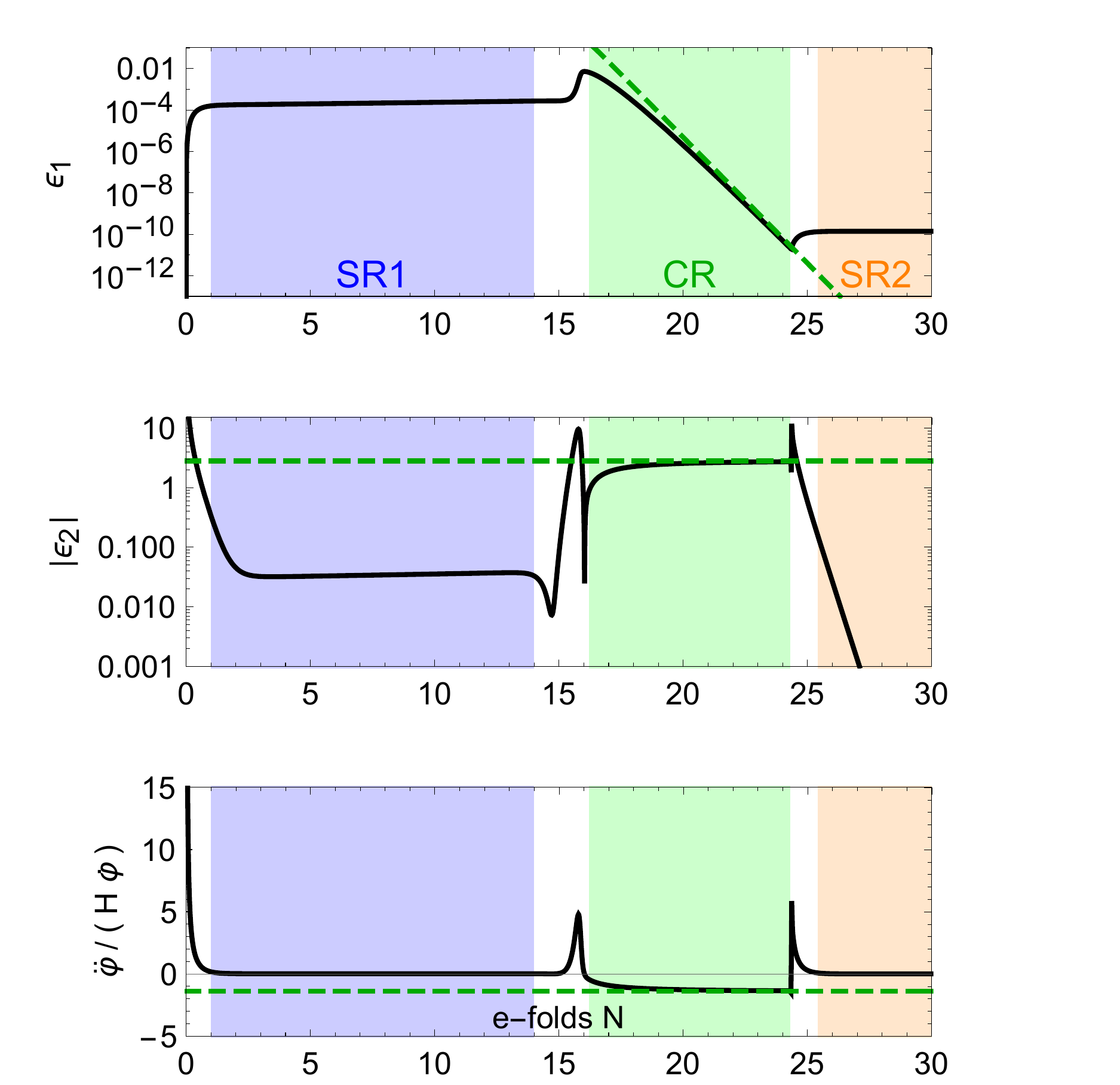}
\caption{
Numerical evolution (\textit{black}) of the first two SR parameters $\epsilon_{1,2}$ and the ratio $\ddot\varphi/(H\dot\varphi)$ for the set of parameters~\eqref{eq:3step_parameters}. The analytical CR asymptotic values are also represented (\textit{dashed, green}). For these background quantities, the transitions last $\mathcal{O}(1)$ $e$-folds. The colored areas correspond to the respective stages of our model (\textit{SR1, CR, SR2}).
}
\label{fig:bg12transition}
\end{figure}

As an example realization, demanding that
\begin{gather}
\beta = -1.4\,,\qquad \frac{m^2}{M^2} = 3.13,\qquad \frac{W_{\textrm{SR2}}}{V_\textrm{CR}' (\varphi_2)} = 5 \,,\nonumber\\
\varphi_s = -5\,,\qquad \varphi_1 = 0.175\qquad \varphi_2 = 4.5\times 10^{-6}\,,\nonumber\\
d_1 = 10^{-2},\qquad d_2 = 10^{-7},
\label{eq:3step_parameters}
\end{gather}
yields the potential and the background evolution shown in Figs.~\ref{fig:pot} and \ref{fig:bg12transition}, respectively. In Fig.~\ref{fig:bg12transition}, while SR stages are simply characterized by $|\epsilon_{1,2}|\ll 1$, the CR stage is recognized by representing the condition \eqref{eq:crcondition}. Each transition lasts for $\mathcal{O}(1)$ $e$-folds. In particular the ratio $m/M$, as well as $\varphi_1$ and $\varphi_2$ were fixed from the requirement of the first transition occurring about \unit[5]{$e$-folds} after the CMB scales exit the horizon to satisfy the observational constraints, and the second transition occurring after sufficient $e$-folds of CR to produce a chosen abundance of PBHs (see Sec.~\ref{sec:3pot} for more details).

% ==================== Figure ====================
% ==================== Figure ====================

% ==================== Figure ====================
\begin{figure}[t]
\centering
\includegraphics[width=8.5cm,angle=0,clip]{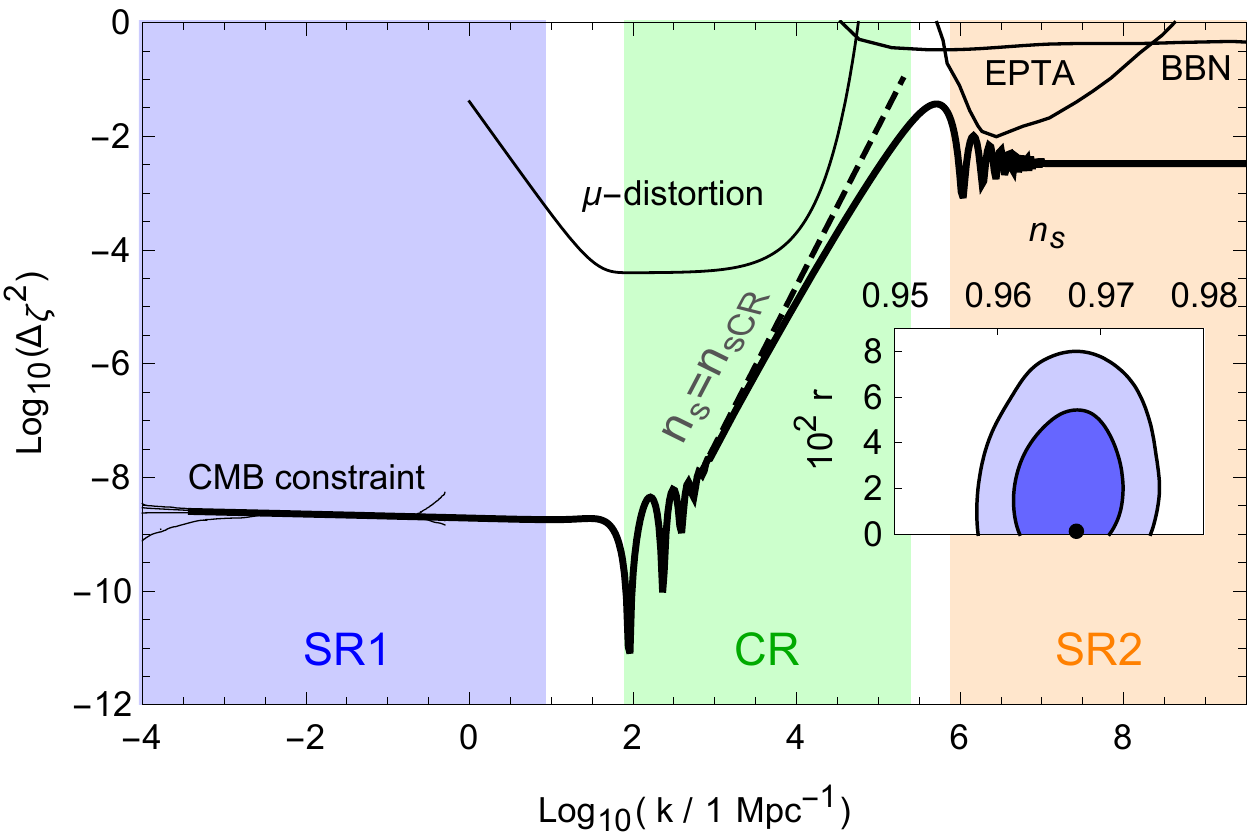}
\caption{Scalar curvature power spectrum for the set of parameters \eqref{eq:3step_parameters} along with the constraint on the scalar tilt and tensor-to-scalar ratio on CMB scales. 
The power spectrum respects the observational constraints such as that from Planck CMB measurements \cite{PlanckCollaboration2018}, $\mu$-distortions (reproduced from \cite{Byrnes2018}), the European Pulsar Timing Array (EPTA) \cite{Lentati2015}, or big bang nucleosynthesis (BBN)-related measurements \cite{Kohri2018}. The latter two constraints are reproduced from \cite{Inomata2018}. Oscillations affect the power spectrum near both transitions, for an effect lasting $\mathcal{O}(1)$ orders of magnitude in $k$. The colored three stages of our model (\textit{SR1, CR, SR2}) are represented according to the horizon crossing scales, i.e.~the scale $k_\textrm{cross}(N) \equiv a(N) H(N)$, where $N$ is chosen in accordance to the same zones as Fig.~\ref{fig:bg12transition}.}
\label{fig:3step_sps}
\end{figure}
% ==================== Figure ====================

Turning to perturbations, the same set of parameters \eqref{eq:3step_parameters} leads to the curvature power spectrum of Fig.~\ref{fig:3step_sps}. 
As for modes that exit the horizon during the first SR stage, scalar and tensor perturbations satisfy the observational constraints.
As expected, the power spectrum on scales that exit the horizon during the CR stage is enhanced up to $\Delta^2_\zeta=\mathcal{O}(10^{-2})$ to produce PBHs.  
The analytically predicted spectral tilt depicted as dashed line in Fig.~\ref{fig:3step_sps} fits the numerical result well.

The transient behavior in the background shown in Fig.~\ref{fig:bg12transition} is reflected to oscillations in the power spectrum in Fig.~\ref{fig:3step_sps}. 
The power spectrum stabilizes to a nearly constant tilt after the oscillatory behavior lasting $\mathcal{O}(1)$ magnitude
orders of $k$. This oscillation can be understood as follows. 
As is shown in Fig.~\ref{fig:bg12transition}, both the transition from SR1 to CR and that from CR to SR2 are in fact characterized by the sharp increase in $\epsilon_1$ at the end of the transient regimes.
The increase causes a time dependence of the mass term $\frac{1}{a^2z}\frac{d^2z}{d\tau^2}$, leading to oscillations in the evolution of the norm of each complex curvature modes that cross the horizon at the transitions.
This leads to an elongated inspiral of the mode in the complex $\zeta$ plane, 
which appears as oscillations in the power spectrum. 

As shown in Fig.~\ref{fig:3step_sps}, 
the power spectrum should satisfy observational constraints on different scales, in particular the most stringent one on CMB scales. 
We consider here the Planck CMB measurements \cite{PlanckCollaboration2018}, which constrain the power spectrum, its tilt and the tensor-to-scalar ratio, the $\mu$-distortion (reproduced from \cite{Byrnes2018}, and note that $y$-distortion will place a constraint on scales between those constrained by CMB and $\mu$-distortion, which does not affect the constraint on scalar tilt in the present case), as well as, along the lines of \cite{Inomata2018}, observations from the European Pulsar Timing Array (EPTA) \cite{Lentati2015}, big bang nucleosynthesis (BBN)-related measurements, and aLIGO measurements. 

One may notice that in Fig.~\ref{fig:3step_sps} the scalar power spectrum remains at $\Delta^2_\zeta\sim 10^{-3}$ on small scales after the peak. 
This originates from the linear approximation~\eqref{VSR2} of the second SR potential $V_\textrm{SR2}(\varphi)$.
As mentioned above, a more realistic choice of $V_\textrm{SR2}(\varphi)$ will transition to a reheating stage, and will lead to a different small-scale power spectrum. However, the PBH production primarily depends on the peak value of the power spectrum on scales that exit the horizon during the CR stage. For this purpose the linear approximation~\eqref{VSR2} of the potential $V_\textrm{SR2}(\varphi)$ is sufficient.

One may also notice, on the latter half of the sharp increase of the power spectrum of Fig.~\ref{fig:3step_sps}, that there is a gradually increasing deviation from the tilt expected for a pure CR case. 
This behaviour can be understood as follows.
In general, curvature modes continue to evolve a few $e$-folds after the horizon exit.
Without the transition from CR to SR2, this evolution would allow the modes continue to grow and reach to the analytical prediction.
In the presence of the transition, this evolution is prevented, which causes the suppression of the power spectrum observed in Fig.~\ref{fig:3step_sps}.
Since the CR to SR2 transition occurs close to the horizon exit on small scales, the amount of would-be enhancement prevented by the transition increases.  
This is the reason why the suppression becomes larger on small scales in Fig.~\ref{fig:3step_sps}.

% ==================== Figure ====================
\begin{figure}[t]
\centering
\includegraphics[width=8.5cm,angle=0,clip]{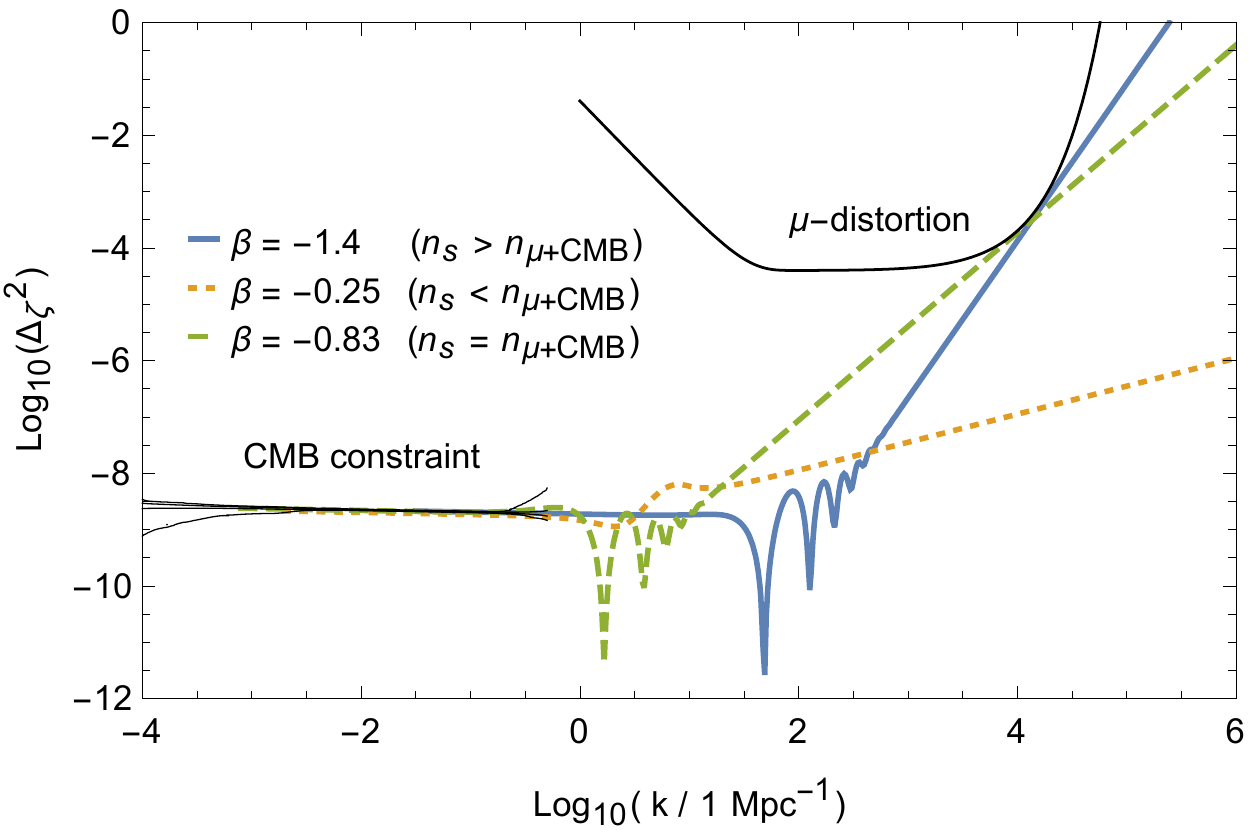}
\caption{
The scalar power spectrum for three different values of $\beta$ 
are impacted differently by the Planck and $\mu$-distortions constraints. The intermediate case $\beta = -0.83$ has both the CMB and the $\mu$-distortion constraints being relevant. Other constraints on smaller scales (not shown here), such as those from gravitational waves, will have an impact for the lower tilt values (see e.g.~Fig.~\ref{fig:3step_sps_4}).
}
\label{fig:triple}
\end{figure}
% ==================== Figure ====================

As highlighted in Sec.~\ref{sec:cr}, the blue tilt of the scalar power spectrum is adjustable for the CR inflation, which is one of the differences from the USR inflation.
Hence, one can obtain different ranges of PBH masses depending on the value of $\beta$.
First, in order to reach the largest PBH masses, one should consider large values of the tilt $n_s$, therefore the limit $\beta\to-\frac{3}{2}$. On the other hand, either a milder tilt or a later transition will both lead to smaller PBH masses (this is also illustrated later in Figs.~\ref{fig:3step_sps_4} and \ref{fig:fPBH}).
Once $\beta$ is chosen, other parameters are adjusted to fit the constraints. Depending on the value of the tilt, different sets of observational constraints will be relevant when calculating the maximum PBH mass (which is simply related to the largest scale at which the amplitude of the perturbations reaches the threshold value $\Delta^2_{\zeta,\textrm{PBH} } \approx 10^{-2}$).

As an example, let us focus on the constraint on the power spectrum from CMB and $\mu$-distortions. Assuming a transition of $\sim1$ order of magnitude in terms of wavenumber (indeed, one cannot simply connect both constraints, due to the oscillations in the curvature power spectrum that occur at the transition), 
we can estimate the critical tilt 
\begin{equation}
n_{\textrm{\tiny$\mu$+CMB}} \approx 2.7\,.
\end{equation}
corresponding to $\beta = -0.83$ at which both constraints (primarily upper bounds on $\varphi_1$) from CMB and $\mu$-distortion are marginally satisfied, as depicted in Fig.~\ref{fig:triple}. 
For tilts larger than $n_{\textrm{\tiny$\mu$+CMB}}$, the principal constraint will be that from $\mu$-distortions. On the other hand for models with $n_s < n_{\textrm{\tiny$\mu$+CMB}}$, the CMB, as well as other small scale constraints (such as that from gravitational waves) will constrain the maximum PBH mass. 
This highlights the fact that the CR inflation accommodates various possibilities of PBH production within the framework of single-field inflation, and that it is robust to future improvements of observational constraints and/or detections.

\section{Primordial black hole production}\label{sec:3pot}%%%%%%%%%%%%%%%%%%%%%%%%%%%%%%%%%%%%%%%% 

In Sec.~\ref{sec:2pot} we presented the three-stage potential~\eqref{eq:pot3step} that transits as SR1 $\to$ CR $\to$ SR2, and highlighted that it allows various blue-tilted scalar spectra with the adjustable scalar spectral tilt.
In this section we connect these results to PBH abundance and compare it with observational constraints.

% ==================== Figure ====================
\begin{figure}[t]
\includegraphics[width=8.5cm,angle=0,clip]{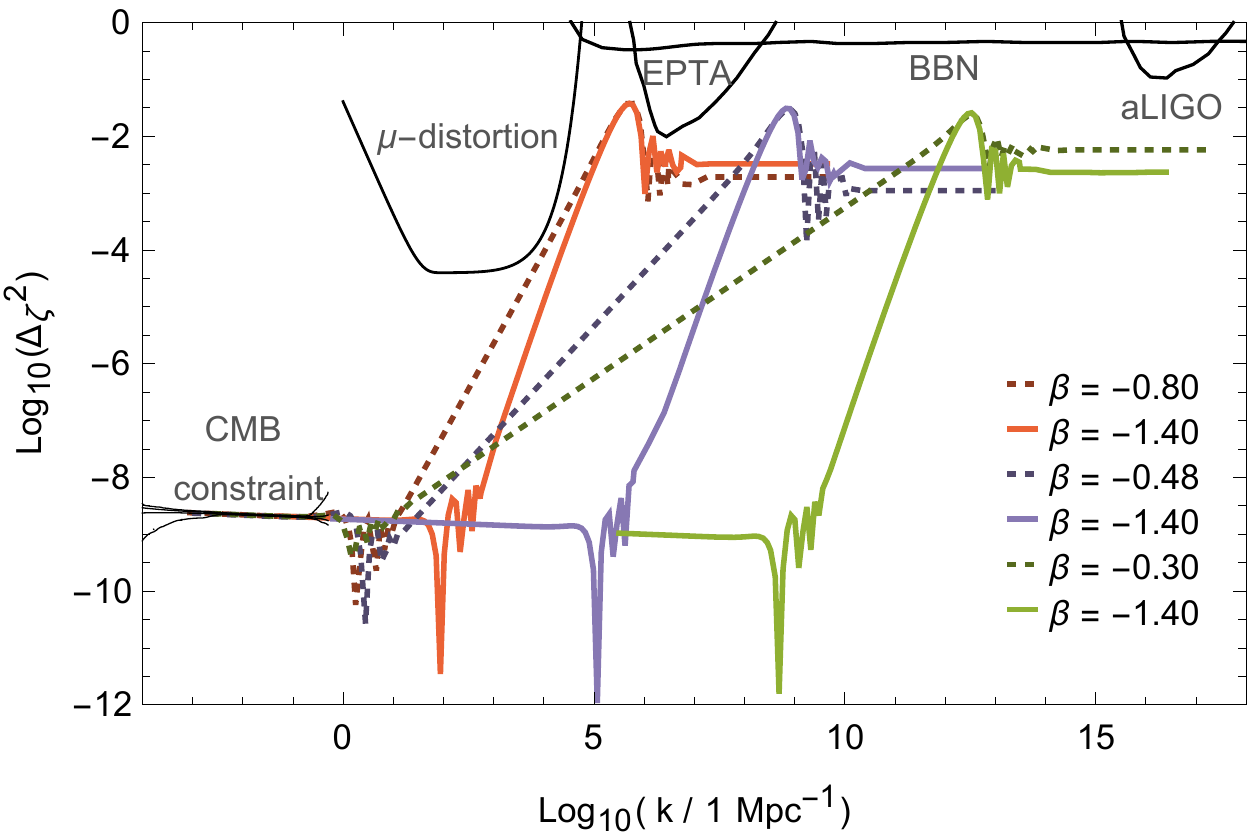}
\caption{Scalar power spectrum for the three-stage potential~\eqref{eq:pot3step}
for values of $\beta$ corresponding to quasi-maximum and quasi-minimum tilts (\textit{solid} and \textit{dashed}, respectively) for each peak.
Modes that exit the horizon during the first SR stage corresponds to the CMB scales, while the intermediate CR stage allows for the power spectrum to increase up to the threshold for PBH production. 
Here the only restriction on the final SR stage is that the slope of the potential increases enough to ensure a slight drop of the power spectrum after it reaches the threshold. 
} 
\label{fig:3step_sps_4}
\end{figure} 
% ==================== Figure ====================

To obtain the PBH abundance, we follow the treatment of \cite{Inomata2017} (and references therein) that models the collapse of an overdense region with some simplifying assumptions. The procedure consists in first evaluating the variance~\cite{Young:2014ana}
\begin{equation}
\sigma^2 (M_\textrm{PBH}(k)) = \frac{16}{81}\int d\ln{q}\, W^2(q/k)\left(q/k\right)^{4} \Delta^2_\zeta(q)\,,
\end{equation}
of the density contrast for the PBH mass of $M_\textrm{PBH}(k)$ coarse-grained by a window function $W(x)$, which we take the Gaussian $W(x)=e^{-x^2/2}$.
It then allows for an estimation within Gaussian statistics of the formation rate
\begin{equation}
\beta_\textrm{PBH}(M_\textrm{PBH}) \simeq \frac{1}{\sqrt{2\pi}}\frac{\sigma(M_\textrm{PBH})}{\delta_c}e^{-\frac{\delta_c^2}{2\sigma^2(M_\textrm{PBH})}}
\end{equation}
where the PBH mass $M_\textrm{PBH}$ is related to the wavenumber $k$ via
\begin{equation}
\frac{M_\textrm{PBH}(k)}{10^{20}\textrm{g} } \simeq \left(\frac{\gamma}{0.2}\right)\left(\frac{g_*}{106.75}\right)^{-\frac{1}{6}}\left(\frac{k}{7\times 10^{12} \textrm{Mpc}^{-1}}\right)^{-2}\,,\label{eq:MPBH}
\end{equation}
in which $g_*$ is the number of relativistic degrees of freedom at PBH formation, and the factor $\gamma$ relates the cosmological horizon mass to the mass of the corresponding PBH. The latter factor depends on the particulars of the formation process and has been the subject of research, see e.g.~\cite{Shibata1999}, but we do not favor any specific value here, since several other parameters (note the large sensitivity on the details of the inflationary background) have still large error margins. Finally, the abundance of PBH over logarithmic mass intervals can be approximated by
\begin{align}\label{eq:PBH_abundnace}
&f_\textrm{PBH}(M_\textrm{PBH}) \equiv \frac{\Omega_\textrm{PBH}(M_\textrm{PBH})}{\Omega_c} \\
&= \left(\frac{\beta(M_\textrm{PBH})}{8\times10^{-15}}\right)\left(\frac{0.12}{\Omega_c h^2}\right)\left(\frac{\gamma}{0.2}\right)^{\frac{3}{2}} 
\left(\frac{106.75}{g_*}\right)^{\frac14}\left(\frac{M_\textrm{PBH}}{10^{20}g}\right)^{-\frac13}\,. \notag
\end{align}
Again, the calculations here can only be taken as rough approximations.
Some approximations taken here (and thus partly in \cite{Inomata2017}) are that of spherical symmetry, constant PBH mass after formation neglecting accretion and merger, and $g_*$ being almost equal to $g_{*s}$.
However, these estimations are sufficient to show the effectiveness of the PBH production in the CR model.

% ==================== Figure ====================
\begin{figure}[t]
\includegraphics[height=5.7cm,angle=0,clip]{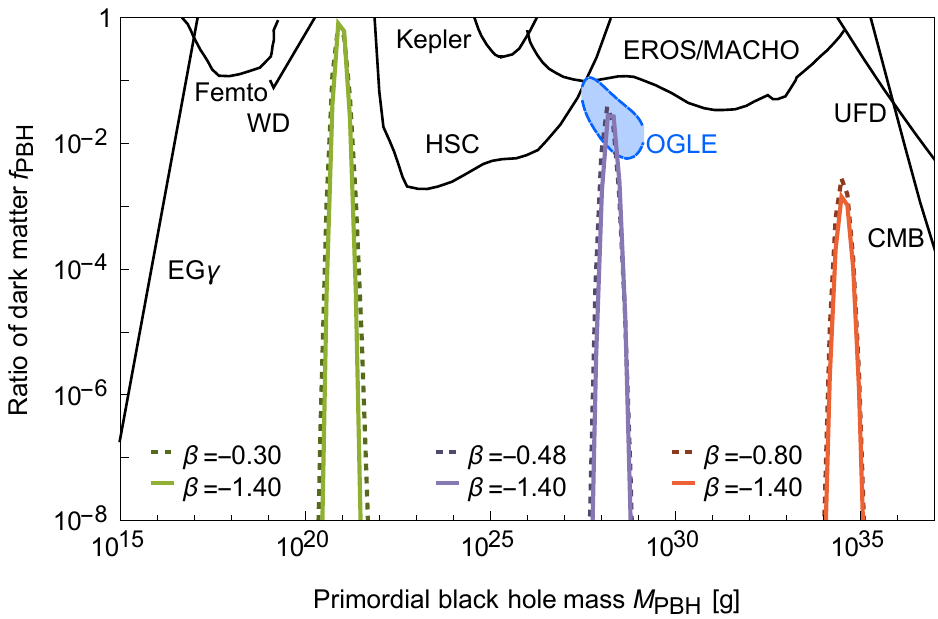}
\caption{PBH abundance for several values of $\beta$, with the same parameters as in Fig.~\ref{fig:3step_sps_4}. Diverse observational constraints are represented, including that on extra-galactic radiation (EG$\gamma$) \cite{Carr:2009jm}, femto-lensing \cite{Ricotti2007}, long-livedness of white dwarfs (WD) \cite{Graham2015}, microlensing by the Subaru HSC \cite{Niikura2017}, Kepler \cite{Griest2013} and by EROS-2 and previous related searches \cite{Tisserand2006}, the survival of ultra-faint dwarf (UFD) galaxies \cite{Brandt2016}, and the accretion on the CMB \cite{Horowitz2016,Poulin2017,Aloni2016,Ali-Haimoud2016}. We also include the allowed parameter space assuming a detection by OGLE \cite{Niikura2019}. Note that these constraints are constantly updated and improved: for example, the femto-lensing and WD constraints have been contested in \cite{Montero-Camacho:2019jte}, and a further constraint in the same range of masses \cite{Laha:2019ssq} has also been proposed. As an other example it has recently been proposed that the HSC constraint be reconsidered \cite{Smyth:2019whb}.}
\label{fig:fPBH}
\end{figure} 
% ==================== Figure ====================

Using the above relations, we convert various blue-tilted scalar power spectra shown in Fig.~\ref{fig:3step_sps_4} to the PBH abundances depicted in Fig.~\ref{fig:fPBH}. 
Exact values of parameters characterizing the potential \eqref{eq:pot3step} employed for Figs.~\ref{fig:3step_sps_4} and \ref{fig:fPBH} are listed in Appendix~\ref{app:params}.
We set other parameters as the same proxy values (e.g. $\gamma = 0.2$, etc.) as given in \eqref{eq:MPBH} and \eqref{eq:PBH_abundnace}.

The possibility to adjust the tilt of the power spectrum renders the CR model robust to observations. Although the tilt of the power spectrum may be changed, this only marginally affects the width of the PBH distribution function. In particular for large PBH masses, this is limited by the widest range of tilts allowed by the different constraints on the scalar power spectrum. While a marginal difference may be produced nevertheless, as seen in Fig.~\ref{fig:fPBH}, 
for a given set of the PBH mass and abundance, the CR model allows a range of the scalar tilt.
In Fig.~\ref{fig:3step_sps_4} are shown large tilt with $\beta=-1.4$, which is close to the maximum tilt with $\beta\to -1.5$, and quasi-minimum tilt for each peak.

In Figs.~\ref{fig:3step_sps_4} and \ref{fig:fPBH}, we focused on three different PBH mass scales, each of which corresponds to PBHs as LIGO-Virgo events (red), OGLE events (purple), and all dark matter (green), the first two of which could be possible detections of primordial black holes.
First, the LIGO-Virgo black holes, which, if shown to be PBHs, would necessitate the abundance $f_\textrm{PBH} \sim 10^{-3}$ at about \unit[30]{$M_\odot$} \cite{Sasaki:2016jop}.
In particular, the red solid line corresponds to the example set of parameters \eqref{eq:3step_parameters}. 
Next, OGLE reported a series of observations possibly consistent with PBHs, which would imply a peaked distribution at $f_\textrm{PBH} \sim 10^{-3/2}$ at about \unit[$10^{26}-10^{28}$]{g} \cite{Niikura2019}. We have also chosen to reproduce this value in Fig.~\ref{fig:fPBH}.
Finally, we provided an example of peak corresponding to PBH as all dark matter scenario at the window around \unit[$10^{21}$]{g}.
Since in this case the required PBH mass is lightest among the three examples, the widest range of the scalar tilt is allowed as depicted in Fig.~\ref{fig:3step_sps_4}.
Since for each case the transient CR scenario allows various tilts of the power spectrum, it can be distinguished from the transient USR scenario and be tested by future observations.

Once more, as explained in Sec.~\ref{sec:2pot}, constant values of $\Delta_\zeta^2$ appeared in Fig.~\ref{fig:3step_sps_4} on small scales after the peaks are not important as they originate from the linear potential approximation~\eqref{VSR2} of the second SR stage.
Ultimately, a more realistic potential should be implemented for the second SR stage to realize a smooth transition to a reheating stage, 
which will also change the small-scale behavior of the power spectrum.
However, the modification would not affect the estimation of PBH abundance very much. 
The linear potential approximation is also sufficient to extract the typical effect of the transition on the power spectrum.

\section{Discussion}\label{sec:conclusion}%%%%%%%%%%%%%%%%%%%%%%%%%%%%%%%%%%%%%%%% 

The constant-roll inflation with $\beta\equiv \frac{\ddot\phi}{H\dot\phi} \approx 0.015$ has been known to produce a slightly red-tilted scalar power spectrum compatible with the observational constraints.
In this work, 
we have instead shed light on the CR inflation with a different parameter range, $-\frac32<\beta<0$, for which a blue-tilted curvature power spectrum is generated during the CR attractor stage without superhorizon growth of curvature modes, and 
have investigated how a stage of CR inflation may lead to the production of PBHs. 
Indeed, PBHs may be generated once a (mildly $k$-dependent) threshold value (approximately $\Delta_\zeta^2 \sim 10^{-2}$) for the curvature power spectrum is reached, which can be easily realized in the CR model since the tilt is adjustable within the range of $0 < n_s - 1 < 3$ for the aforementioned parameter range. 

We have constructed a specific potential~\eqref{eq:pot3step} with three stages. Indeed, the CR stage generating the blue-tilted spectrum should be preceded and followed by SR stages with red tilts in order to satisfy the existing constraints on the curvature power spectrum and on the PBH distribution. 
We have implemented these three stages by using a Starobinsky inflation\textemdash as a proxy for a SR stage satisfying the CMB constraints\textemdash in the region $\phi\gtrsim\phi_1$, the CR stage in the region $\phi_2\lesssim\phi\lesssim\phi_1$, and finally a linear potential approximation in the region $\phi\lesssim\phi_2$\textemdash again a proxy for a SR stage 
to prevent overproduction of PBHs.
The matching positions $\phi_1$ and $\phi_2$ are chosen to satisfy various observational constraints and to partially regulate the PBH mass. 
We find that it is possible for the model to satisfy the observational constraints, and also to induce the PBH abundances necessary in case of a detection via e.g.\ the growth of the family of LIGO-Virgo black holes, or the OGLE microlensing events~\cite{Niikura2019}, or all of dark matter for $M_\textrm{PBH}\sim10^{21}\textrm{ g}$. 

PBH production in the context of single-field inflation has already been studied in 
several
realizations, yet our construction has its own particularities. A stage of the CR inflation, contrary to models which rely on a transient USR stage (such as some analyses of the inflection point potential), in particular has several advantages: 
that the curvature modes are frozen on superhorizon scales as in the standard SR inflation\textemdash which reduces the amount of tuning needed for a desired blue-tilted stage\textemdash, 
that one does not need to be apprehensive for a possible ambiguities in the USR model from stochastic effects on the plateau of the potential, 
and that the tilt of the curvature power spectrum can be simply and freely adjusted. 
The first and second features make the CR model theoretically economical, whereas the third one  
implies that the CR model is robust to more precise observations, for instance on the curvature power spectrum. 
Conversely, future observations constraining small-scale power spectrum such as those planned by the next generation ground- and space-borne gravitational wave observatories (see e.g.~\cite{Inomata2018}) will allow
further observational tests of our model. 
We therefore find that the CR model is an interesting new way to modelize PBH production at scales of interest.

Several further questions are left for future works.
First, it would be interesting to investigate non-Gaussianity of the three-stage CR model.
While, a priori, one may expect that it has a negligible impact on the PBH production (as is the case within USR, see \cite{Passaglia2019}), a dedicated study would be appropriate. 
It would be also interesting to replace the transitions in the potential by a multi-field model, such as curvaton scenario, capable of generating the desired power spectrum on every scale of interest.
Finally, while we focused on the parameter range $-\frac32<\beta<0$ throughout the paper, it would be interesting to further explore the possibilities offered by a wider range of CR models.

\acknowledgments%%%%%%%%%%%%%%%%%%%%%%%%%%%%%%%%%%%%%%%% 

This work was supported in part by Japan Society for the Promotion of Science (JSPS) Grants-in-Aid for Scientific Research (KAKENHI) No.\ JP17H06359 (H.M., S.M.), No.\ JP18K13565 (H.M.), No.\ JP17H02890 (S.M.), and by World Premier International Research Center Initiative (WPI), MEXT, Japan (S.M.), and the Japanese Government (MEXT) Scholarship for Research Students (M.O.).

\appendix

\section{Example parameters}\label{app:params}%%%%%%%%%%%%%%%%%%%%%%%%%%%%%%%%%%%%%%%% 

In this appendix, we report the parameters used to obtain the power spectra in Fig.~\ref{fig:3step_sps_4}, and the PBH abundances in Fig.~\ref{fig:fPBH}, up to normalization.
\vspace{1ex}
\begin{center}
\begin{tabular}{c|cccccc}
$M_\textrm{PBH}$[g] & $10^{21}$ & $10^{21}$ & $10^{28.5}$ & $10^{28.5}$ & $10^{34.5}$ & $10^{34.5}$\\[4pt]
$\beta$ &$-\frac{3}{10}$&$-\frac{7}{5}$&$-\frac{101}{40}$&$-\frac{7}{5}$&$-\frac{4}{5}$&$-\frac{7}{5}$\\[4pt]
$\frac{m^2}{M^2}$&$\frac{511}{165}$&$\frac{12940}{4131}$&$\frac{7770}{2497}$&$\frac{12940}{4131}$&$\frac{890}{285}$&$\frac{12940}{4131}$\\[4pt]
$\frac{W_{\textrm{SR2}}}{V_\textrm{CR}' (\varphi_2)}$ & $2$ & $5$& $5$& $5$& $5$& $5$\\[4pt]
$\varphi_s$ & $-\frac{101}{20}$ & $-5$ & $-5$& $-5$& $-5$& $-5$\\[4pt]
$\varphi_1$ & $\frac{177}{1238}$ & $\frac{95}{543}$ & $\frac{95}{543}$& $\frac{115}{596}$& $\frac{95}{543}$& $\frac{95}{543}$\\[4pt]
$\varphi_2$ & \scriptsize $\frac{619}{24988373}$ & \scriptsize $\frac{1}{270396}$ & \scriptsize $\frac{429}{28438807}$& \scriptsize $\frac{68}{16107079}$& \scriptsize $\frac{83}{10546385}$& \scriptsize $\frac{223}{50550681}$\\[4pt]
$d_1$ & $10^{-3}$ & $10^{-2}$ & $10^{-2}$& $10^{-3}$& $10^{-2}$& $10^{-2}$\\[4pt]
$d_1$ & $10^{-7}$ & $10^{-7}$ & $10^{-7}$& $10^{-7}$& $10^{-7}$& $10^{-7}$
\end{tabular}
\end{center}
\vspace{1ex}
Note that these exact values are only intended as an example, and in particular other parameters can yield comparable results. 

\bibliography{cri_pbh}{}

\end{document}